\documentclass[prl,twocolumn,showpacs,floatfix,superscriptaddress,citeautoscript]{revtex4}

\usepackage{graphicx}

\begin{document}

\title{Vortex dynamics in superconducting channels with periodic constrictions}

\author{K. Yu}
\affiliation{Department of Physics, Syracuse University, Syracuse, NY 13244-1130}
\author{M.B.S. Hesselberth}
\affiliation{Kamerlingh Onnes Laboratorium, Leiden University, P.O. Box 9504, 2300 RA Leiden, The Netherlands}
\author{P.H. Kes}
\affiliation{Kamerlingh Onnes Laboratorium, Leiden University, P.O. Box 9504, 2300 RA Leiden, The Netherlands}
\author{B.L.T. Plourde}
\email[]{bplourde@phy.syr.edu}
\affiliation{Department of Physics, Syracuse University, Syracuse, NY 13244-1130}

\date{\today}


\begin{abstract}
Vortices confined to superconducting easy flow 
channels with periodic constrictions exhibit reversible oscillations in the critical current at which vortices begin moving as the external magnetic field is varied. 
This commensurability 
scales with the channel shape and arrangement, although screening effects play an important role. 
For large magnetic fields, some of the vortices become pinned outside of the channels, leading to magnetic hysteresis in the critical current. Some channel configurations also exhibit a dynamical hysteresis in the flux-flow regime near the matching fields.
\end{abstract}

\maketitle

Vortices flowing through nanofabricated easy flow channels in superconducting films provide a useful system for studying the dynamics of interacting particles moving in tailored confining potentials. 
The general problem of interacting particles in confined geometries is important in a variety of physical systems, 
including colloids flowing through microchannels \cite{Koppl2006} and Wigner crystals \cite{Goldman1990} in the presence of constrictions \cite{Piacente2005}. 
With appropriate asymmetries, such tailored potentials can also form model systems for studying ratchet dynamics, with applications ranging from superconducting devices to investigations of biomolecular motors \cite{Plourde2009}. 
The fabrication of weak-pinning channels for guiding vortices through superconducting films at the nanoscale is well established \cite{pruymboom88}. 
Such channels have been employed in a variety of investigations of vortex dynamics at relatively large magnetic fields, typically greater than $10^3\; {\rm Oe}$, including experiments on mode locking \cite{Kokubo2002} and melting in confined geometries \cite{besseling03}. 

Recent advances in nanofabrication have enabled implementations of artificial periodic vortex pinning lattices in superconducting films. 
These are typically produced with arrays of either 
nanoscale holes through the film  \cite{Baert1995, Welp2005} or magnetic dots underneath the film \cite{Villegas2003}. 
Such structures result in a substantial magnetic field-dependence to the critical current, which is related to the threshold force required to cause vortex motion. 
The critical current typically exhibits commensurate behavior with maxima when the magnetic field corresponds to an integer number of vortices per pinning site. For fields away from these matching points, the dynamics of interstitial vortices, which are not located on the strong pinning sites but rather are more weakly confined through interactions with the strongly pinned vortices, 
lead to lower critical currents. 
A variety of experiments have been performed on such pinning arrays in recent years, including studies of the pinning-strength dependence \cite{Moshchalkov1998}, quasiperiodic lattices \cite{Kemmler2006, Villegas2006}, and structures with random dilutions of pinning sites \cite{Kemmler2009}. There have been many simulations of vortex dynamics in these periodic pinning systems as well \cite{Reichhardt1997, Reichhardt1997b, misko06}. 

\begin{figure}
\centering
  \includegraphics[width=3.35in]{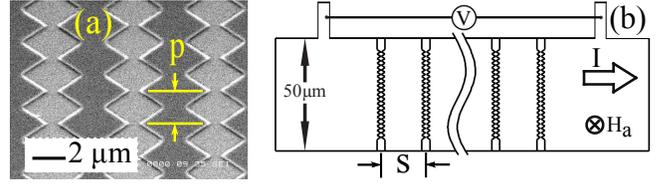}
  \caption{(Color online) (a) Scanning electron micrograph of three channels with periodic diamond constrictions. (b) Strip layout, along with channel and magnetic field orientation. 
\label{fig:schematic}}
\end{figure}

In this article, we describe measurements of vortex dynamics in 
weak-pinning channels that contain periodic constrictions at small magnetic fields, generally less than $10\;{\rm Oe}$. Thus, this involves  considerably smaller fields than much of the previous work on vortex matter in unstructured weak-pinning channels. 
The nature of the channels provides pathways for the easy flow of vortices, while the lattice of periodic constrictions results in strong  
matching effects with substantial enhancement of the critical current $I_c$ at certain values of the external magnetic field $H_a$. 
Although we do not image our vortex distributions directly, we can determine that over much of the field-range of our measurements, all of the vortices are confined to the channels, with the dynamics determined solely by the channel geometry, screening currents in the film, and interactions between vortices. 
Thus, in this field regime there is no distinction between pinned and interstitial vortices. 
At larger $H_a$ vortices can enter the regions outside of the channels where they become strongly pinned and do not participate in the flux-flow. 
Instead, these pinned vortices alter the potential for the vortices that are confined to the channels and lead to an irreversibility of $I_c(H_a)$. 
Besides the magnetic-field hysteresis in $I_c$ for large $H_a$, we often observe a dynamical hysteresis in the vicinity of the matching fields in the current-voltage characteristics (IVCs) themselves.

Following the scheme in Refs.~\cite{pruymboom88, Kokubo2002, besseling03}, we fabricate our channels from bilayer films of a $200\:{\rm nm}$-thick layer of amorphous-NbGe, an extremely weak-pinning superconductor ($T_c^{\rm NbGe} = 2.93$~K), 
and a $50\;{\rm nm}$-thick NbN layer, with relatively strong pinning ($T_c^{\rm NbN} \approx 10$~K), 
on a Si substrate. 
The channels are defined with electron-beam lithography, followed by a reactive ion etching process to remove the NbN, 
resulting in weak-pinning channels for vortices to move through easily. 
The channels are arranged across a $50\,\mu$m-wide strip, with $H_a$ oriented along the thin axis of the strip (Fig. \ref{fig:schematic}). 
The strip pattern contain pairs of probes for coupling to a room-temperature low-noise amplifier for sensing the voltage drop $V$ along the strip due to vortex motion through the channels. 
A transport current driven through the strip with an external supply generates a transverse Lorentz force on the vortices.
Between each pair of voltage probes is an array of $15$ identical channels with inter-channel spacing $s$. 
Each channel contains a periodic chain of cells defined by diamond-shaped constrictions, all of which are $3.2\,\mu$m across at the widest point and $700\,$nm wide at the constriction, with a period along the channel $p$. 
We have measured sets of such channels with five different combinations of $(s, p)$. 

We perform our measurements with the strip immersed in a pumped helium bath. 
Our results presented here were obtained at temperature $T$ between $2.61$~K and $2.90$~K ($89\% - 99\%$ of $T_c^{\rm NbGe}$). 
We can apply the standard dirty-limit expressions to estimate the relevant superconducting parameters of the a-NbGe and NbN films. For the a-NbGe, the coherence length $\xi$ varies between $20 - 80$~nm over the range of $T$, thus, the vortex core size is always much less than the smallest dimension of the channels and the vortex cores are essentially point-like. On the other hand, the penetration depth is quite large, and the thin-film screening length, $\lambda_{\perp} = 2 \lambda^2/d$, where $d$ is the film thickness, ranges between $40 - 370\,\mu$m  for the a-NbGe. 
In the NbN that forms the banks between the channels, $\lambda_{\perp}^{\rm NbN} \approx 8\,\mu$m with little temperature variation since 
$T/T_c^{\rm NbN} \ll 1$ \cite{Yu2007}. Thus, the circulating currents for a vortex in a NbGe channel extend along many, if not all, of the diamond cells in that particular channel and penetrate roughly $8\,\mu$m into the NbN banks on either side of the channel. 
Because $\lambda_{\perp}^{\rm NbGe}$ is much greater than both $\lambda_{\perp}^{\rm NbN}$ and the width of the channels, vortices will be confined to the channels and the shape of the channel walls will play an important role in distorting the circulating currents around each vortex. 
As in our earlier measurements of ratchet dynamics with asymmetrically distorted weak-pinning channels \cite{Yu2007}, by controlling the channel wall shape, it is possible to tailor the confining potential for a vortex in the channel. 

\begin{figure}
\centering
  \includegraphics[width=3.35in]{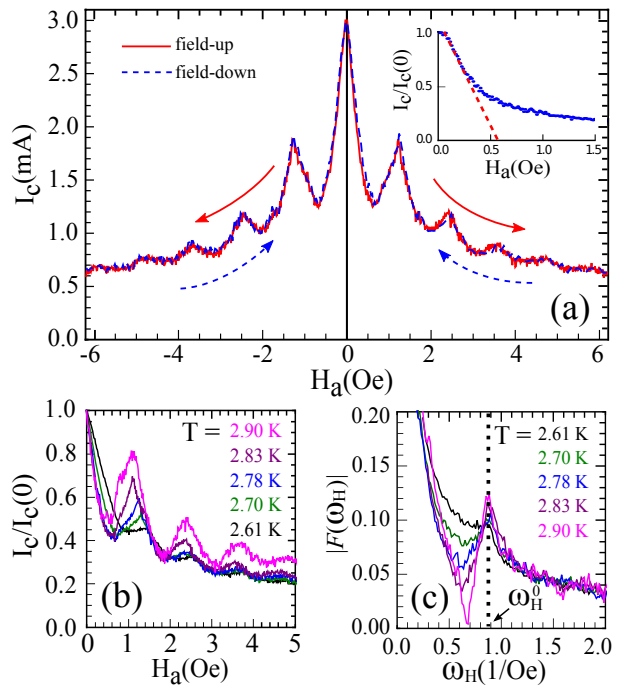}
  \caption{(Color online) (a) Measurement of $I_c(H_a)$ for $p=2\;\mu$m, $s=20\;\mu$m, $T=2.78\;{\rm K}$ for a complete field cycle as described in the text with arrows and legend indicating different portions of magnetic field sweep. 
(inset) $I_c(H_a)$ for $0.5\,\mu$m-wide uniform channels ($s=20\,\mu$m, $T=2.78$ K). (b) $I_c(H_a)$ for same channel parameters as the main figure for different $T$ as indicated, scaled by corresponding $I_c(0)$. (c) Corresponding Fourier transform magnitudes with vertical dotted line indicating location of $\omega_H^0$ for this particular configuration of channels. 
\label{fig:ic-oscillations}}
\end{figure}

We characterize the transition from the static state to a dynamical flux-flow regime by measuring the critical current $I_c$ in the conventional way, that is, by monitoring the current-voltage characteristic, then applying a $1\,\mu$V criterion. 
We drive the vortices with $200$ cycles of a bias current sinusoid at $210$~Hz, then average the resulting voltage response.
We generate $H_a$ with a superconducting Helmholtz coil and a $\mu$-metal shield reduces the background magnetic field below $13$ mG. 
For each measurement sequence, the strip was heated to $\sim 17$~K, well above $T_c$ of both the NbGe and NbN films, and was then cooled in 
$H_a = 0$, while we subsequently increased $H_a$ at the measurement temperature. 

Measurements of the field dependence $I_c(H_a)$ yield information about the vortex dynamics in the channels. 
For comparison, we fabricated a set of 
$0.5$~$\mu$m-wide uniform channels, thus, with no constrictions, and measured $I_c(H_a)$ [Fig. \ref{fig:ic-oscillations}(a)(inset)]. The response is similar to that characteristic of an edge barrier for a thin, weak-pinning superconducting strip in a perpendicular magnetic field, where the entry of vortices at the strip edge is determined by the distortion of the current density across the width of the strip \cite{Plourde2001, benkraouda98}. For a standard edge barrier, $I_c(H_a)$ follows two different regimes: for $H_a$ near zero, $I_c$ decreases linearly with $H_a$, when vortices enter the strip at one edge and are immediately swept across the entire strip width; for larger $H_a$, $I_c \propto H_a^{-1}$, where the external magnetic field is large enough to push vortices into the strip, even for transport currents less than $I_c$. 

The presence of diamond-shaped constrictions in the channels results in pronounced oscillations in $I_c(H_a)$ on top of the edge barrier response [Fig.~\ref{fig:ic-oscillations}(a)]. 
For this measurement, $H_a$ was increased from $0$ to $6.2$ Oe, then reduced through $0$ to $-6.2$ Oe, and finally returned to $0$. The complete reversibility of $I_c(H_a)$ for this field-cycle indicates that all of the vortices are confined to the channels, as one would expect a reversible $I_c(H_a)$ for a pure edge barrier. In contrast, if vortices had entered the strong-pinning NbN, one would expect to observe hysteresis in $I_c(H_a)$. 
The oscillations in $I_c(H_a)$ can be observed over a wide range of $T$ [Fig.~\ref{fig:ic-oscillations}(b)], with the relative height of the peaks increasing as $T/T_c^{\rm NbGe}$ approaches $1$. 
A Fourier transform of the $I_c(H_a)$ data [Fig.~\ref{fig:ic-oscillations}(c)] shows that the characteristic frequency of these oscillations, $\omega_H^0 = 1/\Delta H_a$, as identified by the vertical dotted line in the figure, is independent of $T$ in this range, indicating that the commensurability is determined primarily by the channel geometry. 

\begin{figure}
\centering
  \includegraphics[width=3.35in]{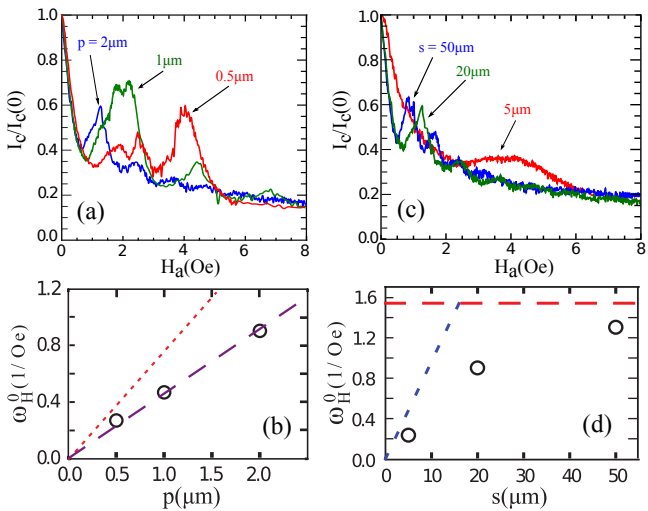}
  \caption{(Color online) (a) $I_c(H_a)$ curves for strips with different $p$ for $s=20\;\mu$m, $T=2.78\;{\rm K}$. (b) $p$-dependence of $\omega_H^0$; dashed line is a guide to the eye (slope$\,=0.46\,$Oe$^{\rm -1}\,\mu$m$^{\rm -1}$); dotted line has slope $2 \lambda_{\perp}^{\rm NbN}/\Phi_0$ ($=0.77\,$Oe$^{\rm -1}\,\mu$m$^{\rm -1}$).  
(c) $I_c(H_a)$ curves for different $s$, with $p=2\;\mu$m, $T=2.78\;{\rm K}$. (d) $s$-dependence of $\omega_H^0$; horizontal dashed line at $2 p \lambda_{\perp}^{\rm NbN}/\Phi_0$, dotted line has slope $p/\Phi_0$. 
\label{fig:s-p-variation}}
\end{figure}

We have studied the commensurability in $I_c(H_a)$ further by measuring a series of channel samples with different values of the diamond cell length $p$ and channel spacing $s$. 
Figure~\ref{fig:s-p-variation}(a) shows $I_c(H_a)$ at $T=2.78$ K for $p=0.5,\,1,\,2\,\mu$m, where all three sets of channels had $s=20\;\mu$m. 
For smaller $p$, the dominant peaks in $I_c$ shift to larger $H_a$, although more complex oscillation patterns develop as well. 
Nonetheless, the Fourier transforms of the $I_c(H_a)$ data indicate that the lowest characteristic frequency in the spectrum for each $p$, $\omega_H^0$, varies linearly with $p$ [Fig.~\ref{fig:s-p-variation}(b)]. 
This provides evidence that the $I_c(H_a)$ peaks are indeed related to a matching of the vortex distribution to the constriction lattice. 
Because each vortex corresponds to one $\Phi_0$ of flux ($\Phi_0 \equiv h c / 2e \approx 20.7 \times 10^{-8}\,{\rm G-cm}^2$), the change in flux density in the channel $\Delta B_{ch}$ that is required to add one vortex to each diamond cell will be determined by the area occupied by this flux. For widely separated channels ($s \gg \lambda_{\perp}^{\rm NbN}$), the flux will extend $\sim \lambda_{\perp}^{\rm NbN}$ into the banks on either side of the channel, while along the channel, the relevant length for the flux is $p$. Thus, one arrives at a rough estimate, $\Delta B_{ch} \approx \Phi_0/2 p \lambda_{\perp}^{\rm NbN}$. However, if $s$ is not large compared to $\lambda_{\perp}^{\rm NbN}$, the resulting overlap between vortices in adjacent channels will lead to an underestimate of $\Delta B_{ch}$. In Fig.~\ref{fig:s-p-variation}(b) we see that a line through the $\omega_H^0(p)$ values has a slope that is approximately $0.6 \times 2 \lambda_{\perp}^{\rm NbN}/ \Phi_0$, 
thus a somewhat larger $\Delta H_a$ is required to achieve a particular $\Delta B_{ch}$. This is likely due in part to neglecting the overlap between vortices ($s=20\,\mu$m in this case), but is also related to the edge barrier mechanism. For a superconducting strip geometry in a perpendicular field, $B$ will be somewhat smaller than $H_a$ due to screening effects until $H_a \gg H_s$, where $H_s$ is the surface entry field \cite{Benkraouda1996a}.

In the opposite limit, $s \ll \lambda_{\perp}^{\rm NbN}$, vortices in adjacent channels will be highly overlapping and the flux density required for a one-vortex change becomes $\Delta B_{ch} \approx \Phi_0/s\,p$. We have 
varied the channel spacing $s$ and observed the influence on $I_c(H_a)$, using $s=5,\,20,\,50\,\mu$m with $p=2\,\mu$m and $T=2.78$ K for all three sets [Fig.~\ref{fig:s-p-variation}(c)]. 
The peak structure shifts to larger $H_a$ for smaller $s$, and the plot of $\omega_H^0$ vs. $s$ in Fig.~\ref{fig:s-p-variation}(d) follows the trends described above, indicated by the dashed and dotted lines included in the plot. 
The $s=50\,\mu$m data approaches the expected $\omega_H^0$ for widely separated channels, while the $s=5\,\mu$m data is close to the limit of highly overlapping vortices. In both cases, one expects a reduction in $\omega_H^0$ somewhat below $\Delta B_{ch}^{-1}$ because of the edge barrier. A detailed calculation of the flux distribution in the channels, accounting for the channel structure, the two different superconductors, and the strip geometry, is beyond the scope of this paper.

\begin{figure}
\centering
  \includegraphics[]{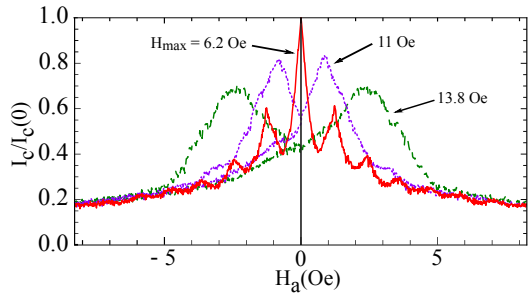}
  \caption{Magnetic hysteresis in $I_c(H_a)$ for larger field sweeps, with $H_{max}$ as indicated for $s=20\;\mu$m, $p=2\;\mu$m, $T=2.78$~K. Curve for $H_{max}=6.2\,$Oe is the same as in Fig. \ref{fig:ic-oscillations} with no hysteresis. 
\label{fig:ha-hysteresis}}
\end{figure}

At the edge of a superconducting strip, vortices will enter when $H_a$ reaches $H_s$, corresponding to the 
current density at the edge reaching a critical level, typically of the order of the Ginzburg-Landau depairing current density. Applying the standard edge barrier expression for $H_s$ \cite{Plourde2001} with our estimated film parameters leads to $H_s^{\rm NbGe} \sim 2.6 - 0.7\,$Oe, although the entry field into the ends of the NbGe channels is likely somewhat smaller than the $H_s^{\rm NbGe}$ estimate when one accounts for current distortions at the channel ends. Indeed, we typically observe the first entry of vortices into the channels followed by oscillations in $I_c(H_a)$ for $H_a \sim 1\;{\rm Oe}$. Performing a similar estimate for vortex entry into the NbN banks yields $H_s^{\rm NbN} \sim 8\,$Oe. 
We can probe the possibility of vortex entry into the NbN by increasing $H_a$ to progressively larger values $H_{max}$ before reducing it and checking the reversibility of $I_c(H_a)$, as vortices trapped in the strong-pinning NbN will exhibit an irreversible magnetic response and will offset the net magnetic field experienced by the vortices confined to the channels. For small $H_{max}$, $I_c(H_a)$ retraces completely [Fig. \ref{fig:ic-oscillations}(a)], corresponding to the entry of vortices only into the NbGe channels. However, for $H_{max} \gtrsim 8\,$Oe, $I_c(H_a)$ becomes hysteretic, with the opening of the hysteresis loop growing with $H_{max}$ (Fig. \ref{fig:ha-hysteresis}). 
Also, the matching peak structure on the return branches of $I_c(H_a)$ becomes washed out for larger $H_{max}$, as the disordered distribution of vortices that occurs in the strong-pinning NbN when $H_a$ is reduced randomizes the potential for the vortices moving in the channels. 

\begin{figure}
\centering
 \includegraphics[width=3.35in]{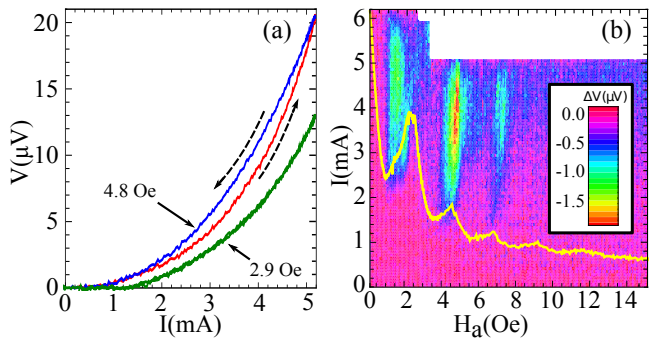}
 \caption{(Color online) (a) Example of dynamical hysteresis in IV curve in the vicinity of one of the matching peaks (red/blue) compared to a reversible IV curve (green) [$s=20\;\mu$m, $p=1\;\mu$m, $T=2.70\;{\rm K}$]. (b) Density plot of the difference of the flux-flow voltage between the outgoing and return current sweeps [$\Delta V = V_{out}(I) - V_{return}(I)$] as the color scale for different $H_a$; corresponding $I_c(H_a)$ superimposed (yellow). 
\label{fig:iv-hysteresis}}
\end{figure}

For $H_a$ below the threshold to introduce vortices into the NbN banks, 
in the vicinity of the $I_c(H_a)$ matching peaks, we often observe a completely different type of dynamical irreversibility consisting of hysteresis in the IV curves. Figure~\ref{fig:iv-hysteresis}(a) shows two example IVCs for the outgoing and return current sweeps, one between matching peaks with no hysteresis, the other near the second matching peak with clear hysteresis. 
Thus, in this second example, there is a clear irreversibility in the flux-flow voltage, but not in the critical current. 

We can combine all of the measured IVCs for a particular channel configuration and $T$ by making a density plot, where the color scale is the difference between the voltage on the outgoing and return current sweeps. We superimpose the corresponding $I_c(H_a)$ curve for reference [Figure~\ref{fig:iv-hysteresis}(b)]. This particular example, with $s=20\;\mu$m, $p=1\;\mu$m, $T=2.70\;{\rm K}$, shows regions of dynamical hysteresis near the first three $I_c(H_a)$ peaks. 
Over the range of drive frequency that we have studied, $20 - 400$ Hz, we observe no change in this response. 
The upper limit of the data on the current axis is set by the point where the flux-flow voltage approaches the 
Larkin-Ovchinnikov instability point \cite{larkin75}, where the channels switch abruptly to the normal state. 

\begin{figure}
\centering
 \includegraphics[width=3.35in]{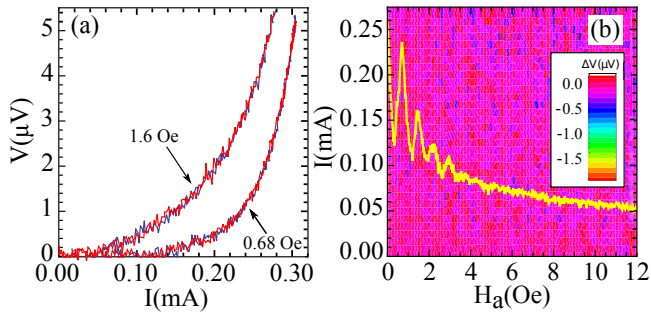}
 \caption{(Color online) Example of measurement at higher temperature with no dynamical hysteresis present [$s=50\;\mu$m, $p=2\;\mu$m, $T=2.90\;{\rm K}$]: (a) sample IV curves for outgoing and return sweeps together, one in vicinity of one of the matching peaks ($0.68\,$Oe) and one in between matching peaks ($1.6\,$Oe). (b) Density plot for this channel configuration and temperature calculated with same technique as in Fig. \ref{fig:iv-hysteresis}(b); corresponding $I_c(H_a)$ superimposed (yellow). 
\label{fig:iv-nohysteresis}}
\end{figure}

This hysteresis in the IVCs may correspond to a distortion of the vortex distribution as the driving current is reduced that allows the vortices to keep flowing at higher velocities than when the current was initially increased. 
We note that not all of the diamond channel measurements displayed this dynamical hysteresis.
In particular, none of the channel configurations that we have studied exhibited this type of irreversibility when measured at the highest temperature of our experiments, $T=2.90$ K. 
Figure~\ref{fig:iv-nohysteresis} contains example IVCs and a similar density plot with superimposed $I_c(H_a)$ curve to that in Fig.~\ref{fig:iv-hysteresis}(b) for a set of channels with $s=50\;\mu$m, $p=2\;\mu$m, measured at $T=2.90\;{\rm K}$ where there is no evidence of dynamical hysteresis. 
This may be due to the change in the intervortex interaction strength as $T$ approaches $T_c^{\rm NbGe}$. 
We are currently investigating this dynamical hysteresis in our channels further. 
We note that hysteretic dynamics for vortices in periodic arrays of antidots were recently reported \cite{gutierrez09}. These were connected to previous theoretical work involving the transition to turbulent flow related to the interplay between interstitial vortices and those pinned in the antidots \cite{Reichhardt1997,misko06}. The origin of the hysteresis in our system is likely somewhat different, as all of the vortices are confined to the weak-pinning channels. 

In summary, we have measured vortex dynamics in weak-pinning channels containing periodic constrictions that are small compared to the vortex size. Over much of the magnetic field range that we have studied, all of the vortices are confined to the channels and the channel structure results in strong matching effects between the vortex distribution and the constriction lattice. In the vicinity of the matching peaks, we often observe a dynamical hysteresis in the vortex response that may be related to a distortion of the vortex distribution. 

We acknowledge useful discussions with J. Clem, A. Middleton, and V. Misko. 
This work was supported by the National Science Foundation under Grant DMR-0547147. We acknowledge use of the Cornell NanoScale Facility, a member of the National Nanotechnology Infrastructure Network, which is supported by the National Science Foundation (Grant ECS-0335765).

\bibliography{vortex}

\end{document}